\documentclass{article}
\usepackage{spconf,amsmath,graphicx,hyperref}
\usepackage{multirow}
\usepackage{multicol}
\usepackage{siunitx}
\usepackage{enumitem}
\usepackage{amssymb}
\usepackage{comment}
\usepackage{float}
\usepackage[table]{xcolor}
\usepackage{url}

\title{Do Foundational Audio Encoders Understand Music Structure?}
\name{
Keisuke Toyama$^{1*\dag}$\thanks{*~These authors contributed equally to this study.}\thanks{\dag~We would like to thank Marco A. Mart\'{i}nez-Ram\'{i}rez for his valuable comments while preparing this manuscript.}\;
Zhi Zhong$^{1*}$\;
Akira Takahashi$^{1}$\;
Shusuke Takahashi$^{1}$\;
Yuki Mitsufuji$^{1,2}$
}
\address{
$^{1}$Sony Group Corporation, Japan \;
$^{2}$Sony AI, USA\
}
\begin{document}
\ninept
\maketitle
%
\begin{abstract}
In music information retrieval (MIR) research, the use of pretrained foundational audio encoders (FAEs) has recently become a trend.
FAEs pretrained on large amounts of music and audio data have been shown to improve performance on MIR tasks such as music tagging and automatic music transcription.
However, their use for music structure analysis (MSA) remains underexplored: only a small subset of FAEs has been examined for MSA, and the impact of factors such as learning methods, training data, and model context length on MSA performance remains unclear.
In this study, we conduct comprehensive experiments on 11 types of FAEs to investigate how these factors affect MSA performance.
Our results demonstrate that FAEs using self-supervised learning with masked language modeling on music data are particularly effective for MSA.
These findings pave the way for future research in FAE and MSA.
\end{abstract}

\vspace{-2mm}
\section{Introduction}
\label{sec:sec1_introduction}
\vspace{-1mm}
Music structure analysis (MSA) is a fundamental and challenging task in music information retrieval (MIR)~\cite{MSA_2020_nieto}.
MSA aims to divide music recordings into non-overlapping consecutive segments and assigning labels that define their functions, such as `intro,' `verse,' `chorus,' and `bridge'~\cite{MSA_2020_nieto, MSA_2011_smith}.
This process typically involves two main subtasks: \textit{boundary detection}, which identifies the temporal points where segments are separated, and \textit{function prediction}, which classifies segments based on musical similarity and features~\cite{MSA_2020_nieto}.

MSA algorithms leverage key musical characteristics: segments are relatively homogeneous in attributes like key and instrumentation; sudden shifts in musicality occur at segment transitions; and segments with the same functional label are musically similar and often repeat~\cite{MSA_2020_nieto}.
Using these characteristics, methods based on self-similarity matrices of audio features such as mel-frequency cepstral coefficients (MFCCs)~\cite{MSA_2019_tralie,MSA_2021_salamon,MSA_2023_buisson} and chromagrams~\cite{MSA_2019_tralie,MSA_2021_bytedance_ismir,MSA_2023_buisson,MSA_2024_chen} have been proposed.
With the development of annotated datasets such as SALAMI~\cite{MSA_2011_smith} and Harmonix~\cite{dataset_harmonix}, supervised learning methods have emerged, particularly those using deep neural networks (DNNs) such as convolutional neural networks (CNNs)~\cite{MSA_2014_ullrich,MSA_2021_bytedance_icassp}, Transformers~\cite{MSA_2022_bytedance_icassp,MSA_2023_kaist,MSA_2025_bytedance}, and approaches combining self-similarity with other features~\cite{MSA_2019_tralie,MSA_2021_bytedance_ismir,MSA_2021_salamon,MSA_2022_peeters,MSA_2023_buisson,MSA_2023_peeters,MSA_2024_chen}.
The majority of conventional methods rely on audio features and neural network architectures that are both hand-crafted and task-specific.

Advances in representation learning have led to various pretrained models that extract general-purpose features from audio and music~\cite{AE_MusicFM,AE_MERT,AE_MetaAudioMAE,AE_SonyAudioMAE,AE_MULE,AE_PANNs,AE_PaSST,AE_CLAP,AE_OpenL3}, offering task-independent alternatives to various MIR tasks.
In this study, we refer to these models as \textit{foundational audio encoders} (FAEs).  
Previous studies have shown that FAE features improve MIR tasks, including music tagging~\cite{downstream_2022_jukebox,AE_MERT}, piano transcription~\cite{downstream_2025_sonido}, music source separation~\cite{downstream_2025_sonido}, and music super-resolution~\cite{AE_SonyAudioMAE}.
Recently, two studies have applied this approach to MSA.  
Won et al.~\cite{AE_MusicFM} investigated MusicFM~\cite{AE_MusicFM} and MERT~\cite{AE_MERT}, finding that FAEs can outperform conventional methods with a simple two-layer linear backend. 
Zhang et al.~\cite{MSA_2025_bytedance} improved performance by connecting MusicFM to a Transformer backend that captures a longer context of the FAE features.

Although many open-source FAEs have been proposed, only two have been examined in MSA research.  
While FAEs differ in learning objectives, methods, training data types, and neural network architectures, how these factors contribute to MSA performance remains unclear.  
Additionally, several FAEs have been used as evaluation metrics for audio and music generation tasks~\cite{AE_MERT,AE_PANNs,AE_PaSST,AE_CLAP,AE_OpenL3,AE_EnCodec}, but their effectiveness for MSA is not well understood.
Finding an FAE that understands music structure could lead us to a better metric for long-form or full-track music generation. 
Surveys and benchmarks have explored using Fr\'{e}chet Audio Distance (FAD)~\cite{metrics_fad} and other combinations of distance metrics and FAEs to evaluate generated music~\cite{review_fad,review_mad}, and as codec benchmarks~\cite{review_espnet}, but such research for MSA tasks is still lacking.

Our contributions are as follows.
First, we present a comprehensive review of 11 FAEs (Table \ref{tab:table1_overview}), considering potential design factors that may affect their performance on MSA.
Second, we visualize FAE features (Fig. \ref{fig:fig1_FAE_feature}) to understand the different behaviors of FAEs qualitatively.
Finally, we conduct linear probing on various FAEs (Table \ref{tab:table2_result_8fold}).
Our results demonstrate that FAEs trained with self-supervised masked language modeling on music data are particularly effective for MSA.
Our code is publicly available.\footnote{\url{https://github.com/sony/MSA-bench}}

\vspace{-2mm}
\section{Foundational Audio Encoder}
\label{sec:sec2_foundational_audio_encoder}
\vspace{-1mm}
{
\begin{table*}[t]
\centering
\footnotesize
\sisetup{
    reset-text-series = false, 
    text-series-to-math = true, 
    mode=text,
    tight-spacing=true,
    round-mode=places,
    round-precision=1,
    table-format=2.1,
    table-number-alignment=center
}
\begin{tabular}{c|cccccccc}
\multirow{2}{*}{FAE}&\multirow{2}{*}{Architecture}&\multirow{2}{*}{\# Param}&\multicolumn{2}{c}{Training data}&\multirow{2}{*}{Fs}&\multirow{2}{*}{Frame rate}&\multirow{2}{*}{Feature dim}&\multirow{2}{*}{Layer}\\
&&&Dataset&Length&&&\\
\hline
\rowcolor{gray!20}\multicolumn{9}{l}{\it Self-supervised Learning: Masked Language Modeling (MLM)}\\
MusicFM&Transformer&330M&Music (FMA~\cite{dataset_fma}, MSD~\cite{dataset_msd})&30 s&24 kHz&25 Hz&1024&8th\\
MERT (95M)&Transformer&95M&Music (1k-hour private)&5 s&24 kHz&75 Hz&768&8th\\
MERT (330M)&Transformer&330M&Music (160k-hour private)&5 s&24 kHz&75 Hz&1024&14th\\
AudioMAE (Huang)&Transformer&85M&Audio (AudioSet~\cite{dataset_audioset})&10 s&16 kHz&6.25 Hz&768&12th\\
AudioMAE (Zhong)&Transformer&85M&Music (private)&5 s&16 kHz&6.25 Hz&3840&12th\\
\hline
\rowcolor{gray!20}\multicolumn{9}{l}{\it Self-supervised Learning: Contrastive Learning}\\
MULE&CNN&-&Music (private)&3 s&16 kHz&0.5 Hz&1728&-\\
\hline
\rowcolor{gray!20}\multicolumn{9}{l}{\it Self-supervised Learning: Tokenization (Codec)}\\
EnCodec (24 kHz)&CNN&7M&Audio (mixture of public data)&1 s&24 kHz&75 Hz&128&-\\
EnCodec (48 kHz)&CNN&7M&Music (MTG-Jamendo~\cite{dataset_mtg_jamendo})&1 s&48 kHz&150 Hz&128&-\\
DAC&CNN&22M&Audio (mixture of public data)&0.38 s&44.1 kHz&86 Hz&1024&-\\
\hline
\rowcolor{gray!20}\multicolumn{9}{l}{\it Supervised Fine-tuning (Audio Tagging) after MLM}\\
AudioMAE (Huang)&Transformer&85M&Audio (AudioSet)&10 s&16 kHz&6.25 Hz&768&12th\\
AudioMAE (Zhong)&Transformer&85M&Music (private)&5 s&16 kHz&6.25 Hz&3840&11th\\
\hline
\rowcolor{gray!20}\multicolumn{9}{l}{\it Supervised Learning (Audio Tagging)}\\
PANNs&CNN&80M&Audio (AudioSet)&10 s&32 kHz&0.1 Hz&2048&-\\
PaSST&Transformer&86M&Audio (AudioSet)&10 s&32 kHz&20 Hz&768&*\\
\hline
\rowcolor{gray!20}\multicolumn{9}{l}{\it Supervised Learning \& Fine-tuning (Sound Event Detection)}\\
PANNs&CNN&80M&Audio (AudioSet)&10 s&32 kHz&3.125 Hz&2048&-\\
\hline
\rowcolor{gray!20}\multicolumn{9}{l}{\it Cross-modal Contrastive Learning (Audio-text)}\\
CLAP&Transformer&31M&Audio (mixture of public data)&10 s&48 kHz&0.1 Hz&512&*\\
\hline
\rowcolor{gray!20}\multicolumn{9}{l}{\it Cross-modal Contrastive Learning (Audio-visual)}\\
OpenL3&CNN&4M&Audio (AudioSet)&1 s&48 kHz&1 Hz&6144&-
\end{tabular}
\vspace{-2mm}
\caption{Overview of foundational audio encoders. In the \textit{Dataset} column, \textit{Audio} denotes short audio clips from diverse audio sources (sometimes including music), and \textit{Music} denotes a music dataset containing long-form and/or full-track music.
The \textit{Layer} column indicates the Transformer layers used in the experiments described in Sec.~\ref{sec:sec3_experiments}.
*For PaSST and CLAP, the official API is used without specifying a layer.}
\vspace{-3.5mm}
\label{tab:table1_overview}
\end{table*}
}

Table \ref{tab:table1_overview} provides an overview of the 11 FAEs investigated in this study.
A visualization of their features (Fig. \ref{fig:fig1_FAE_feature}) reveals a visible correlation with the music structure annotations at a macro level.

\vspace{-2.5mm}
\subsection{Self-supervised Learning (SSL)}
\label{sec:sec2_1_self_supervised_learning}
\vspace{-1mm}
\noindent\textbf{Masked Language Modeling (MLM).}
\textit{MusicFM}~\cite{AE_MusicFM}, \textit{MERT}~\cite{AE_MERT}, \textit{AudioMAE (Huang et al.)}~\cite{AE_MetaAudioMAE}, and \textit{AudioMAE (Zhong et al.)}~\cite{AE_SonyAudioMAE} are Transformer-based models trained with MLM.
Except for AudioMAE (Huang), which was trained using AudioSet~\cite{dataset_audioset}, all the others were trained on music data.
MusicFM has a longer context length of 30 seconds, enabling the use of long-term temporal context, whereas the others have a shorter 5-second context length.
The two AudioMAEs differ in their feature dimension and training data (audio vs. music).
Both also have variants fine-tuned for the audio tagging task~\cite{dataset_audioset}.
MERT, a model with a higher frame rate of 75 Hz, is used as a metric to evaluate the quality of music generation, such as FAD~\cite{review_fad} and MAUVE Audio Divergence (MAD)~\cite{review_mad}.
In Fig. \ref{fig:fig1_FAE_feature}, MusicFM, MERT, and AudioMAE (Zhong) exhibit similar fine-grained repeating patterns over the time axis. 

\smallskip
\noindent\textbf{Contrastive Learning.}
\textit{MULE}~\cite{AE_MULE} is a CNN-based model trained with contrastive learning.
It is trained on music data, though the length is relatively short at 3 seconds.
Another significant difference from other SSL models is its coarse frame rate at 0.5 Hz.

\smallskip
\noindent\textbf{Tokenization (Codec).}
\textit{EnCodec}~\cite{AE_EnCodec} and \textit{DAC}~\cite{AE_DAC} are CNN-based codecs capable of compressing input at high compression rates.
The former can compress a 24-kHz and 48-kHz signal to under 24 kbps, while the latter can compress a 44.1-kHz signal to 8 kbps.
The EnCodec 48-kHz model was trained on music data, while the others were trained on various audio data.
Both models operate at very high frame rates, which can be observed in the visualized features in Fig. \ref{fig:fig1_FAE_feature}.
The training data length is very short (under one second), suggesting long-term context is likely unavailable.
Both models are used as a metric for FAD~\cite{review_fad}.

\vspace{-2.5mm}
\subsection{Supervised Learning}
\label{sec:sec2_2_supervised_learning}
\vspace{-1mm}
\textit{PANNs}~\cite{AE_PANNs}, a CNN-based model, and \textit{PaSST}~\cite{AE_PaSST}, a Transformer-based model, are both trained in a supervised manner on the audio tagging task using AudioSet~\cite{dataset_audioset}.
PaSST has a frame rate of 20 Hz, while PANNs provides only a single feature vector per 10-second input.
The above two models are used as a metric to evaluate the quality of audio generative models~\cite{metrics_audioldm,metrics_specmaskfoley}.
PANNs has another model fine-tuned for the sound event detection task~\cite{task_sound_event_detection}, which can obtain features with a frame rate of 3.125 Hz.

\vspace{-2mm}
\subsection{Cross-modal Contrastive Learning}
\label{sec:sec2_3_cross_modal_contrastive_learning}
\vspace{-1mm}
\textit{CLAP}~\cite{AE_CLAP} and \textit{OpenL3}~\cite{AE_OpenL3} are models trained with cross-modal contrastive learning.
CLAP is a Transformer-based model trained on audio-text pairs, while OpenL3 is a CNN-based model trained on audio-video pairs.
Both are trained on diverse audio data, but they differ in input length and output format: CLAP, similar to PANNs, processes 10-second inputs to produce a single feature vector, whereas OpenL3 uses shorter 1-second inputs to generate frame-wise features at a coarse rate of 1 Hz.
CLAP is used as a metric for FAD~\cite{review_fad} and MAD~\cite{review_mad}, while OpenL3 is used for FAD in the Stable Audio series~\cite{metrics_stable_audio_2}, which focuses on long-form music generation.

\vspace{-2mm}
\section{Experiments}
\label{sec:sec3_experiments}
\vspace{-2mm}
\subsection{Dataset and Metrics}
\label{sec:sec3_1_dataset_and_metrics}
\vspace{-2mm}
\smallskip
\noindent\textbf{Dataset.}
We used the Harmonix~\cite{dataset_harmonix} dataset, which consists of 912 songs totaling approximately 3,400 minutes and spans multiple genres.
We converted the function labels in the dataset into seven types (`intro,' `verse,' `chorus,' `bridge,' `inst,' `outro,' and `silence') according to~\cite{MSA_2022_bytedance_icassp}.  
We performed 8-fold cross-validation with a 6-1-1 split for training, validation, and testing, respectively.

\smallskip
\noindent\textbf{Metrics.}
For boundary detection, we used the F-score at a hit rate at 0.5 seconds (HR.5F) and at 3 seconds (HR3F).
For function prediction, we used the F-score for pairwise frame-level clustering (PWF) and the accuracy rate per frame (ACC).
All metrics are widely adopted in MSA~\cite{MSA_2020_nieto}.
We calculated these scores using the \texttt{mir\_eval} library~\cite{mir_eval} with its default settings.

\begin{figure}[t]
    \centering
    \includegraphics[width=0.98\linewidth]{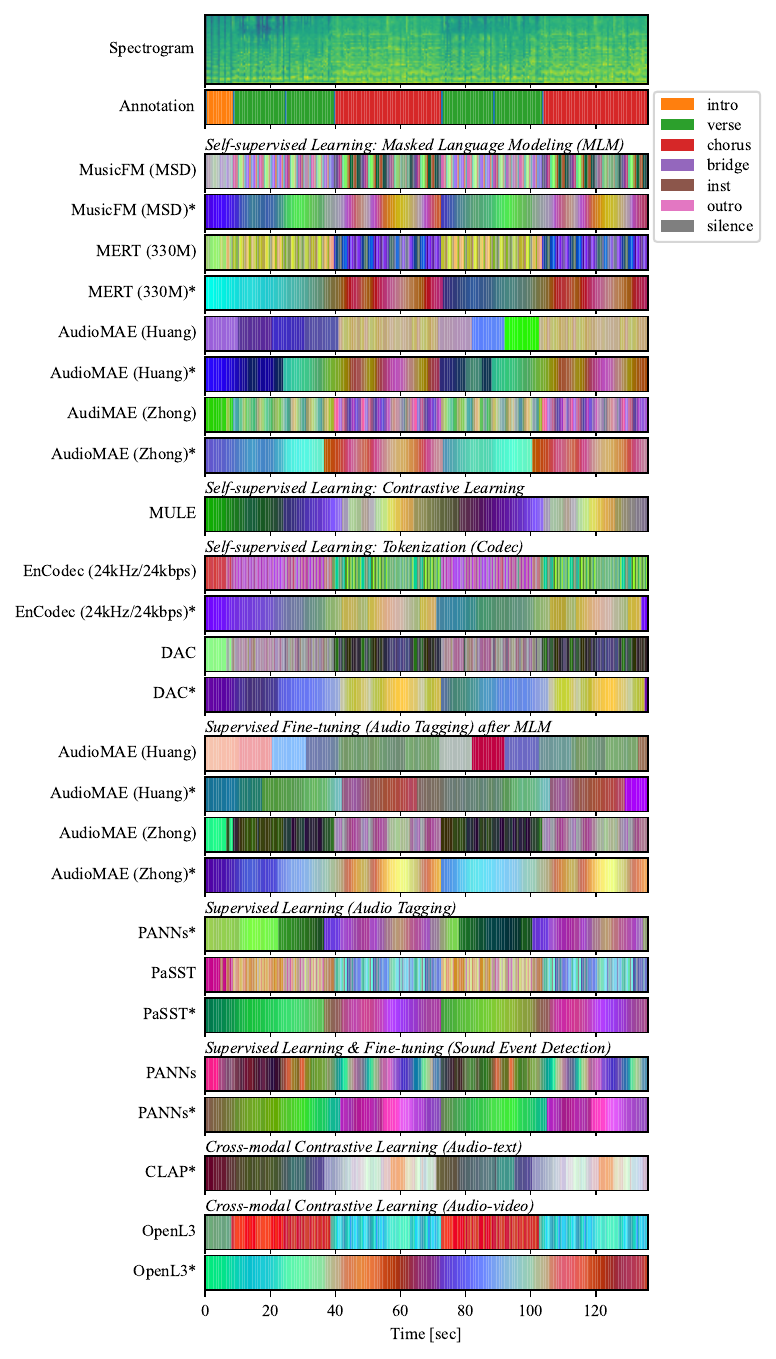}
    \vspace{-5mm}
    \caption{Visualization of FAE features. UMAP~\cite{umap} is used to project embeddings into 3-dimensions, which are then rendered with red, green, and blue colors, respectively. * indicates pooling described in Sec.~\ref{sec:sec3_2_methods}, which smoothes out the original features.}
    \label{fig:fig1_FAE_feature}
    \vspace{-4mm}
\end{figure}

\vspace{-2mm}
\subsection{Methods}
\label{sec:sec3_2_methods}
\vspace{-1mm}

\noindent\textbf{FAE Feature Extraction.}
We extracted features from each FAE following the official implementations and instructions provided in their respective GitHub repositories.

\smallskip
\noindent\textbf{Pooling.}
Since MSA is a task that requires long-term understanding of music, we hypothesize that some information in the fine-grained FAE features is redundant.
Furthermore, since HR.5F, the strictest metric, has a tolerance of 0.5 seconds, a frame rate of approximately 2 Hz may be sufficient.
To investigate this, we performed a \textit{pooling} process.
Feature vectors were obtained from 5 seconds of input and averaged along the time axis to form a single embedding.
Performing this with a 0.5-second hop size generated pseudo-features with a temporal resolution of 2 Hz.
The smoothing effect of pooling can be observed in Fig. \ref{fig:fig1_FAE_feature}.
Since we cannot obtain frame-wise features from PANNs and CLAP, the pooling method allowed us to pseudo-generate frame-wise features.

{
\setlength{\intextsep}{2pt}
\setlength{\floatsep}{2pt}
\begin{figure}[t]
    \centering
    \includegraphics[width=0.95\linewidth]{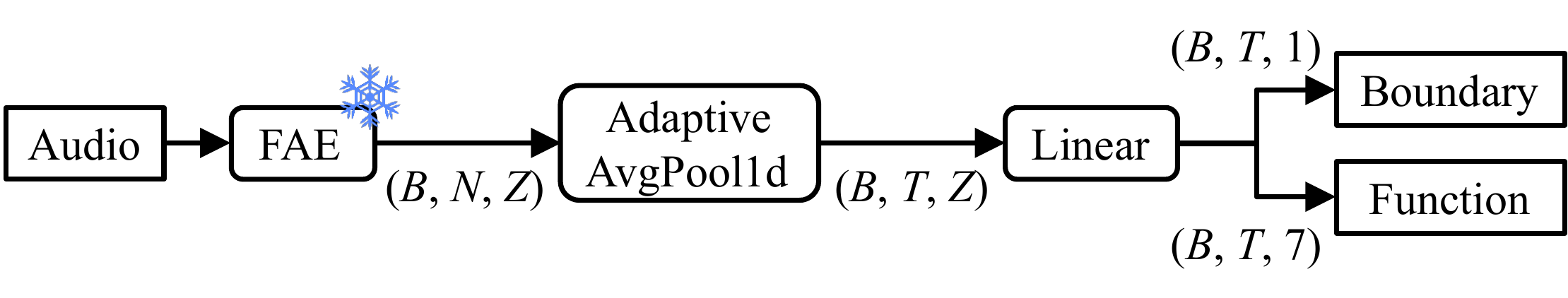}
    \vspace{-5mm}
    \caption{Linear probing. $B$: batch size. $Z$: Feature dimension. $N$ and $T$: number of feature frames before and after adaptive pooling.}
    \vspace{-4mm}
    \label{fig:fig2_linear_pooling_model}
\end{figure}
}

\smallskip
\noindent\textbf{Linear probing model.}
To investigate the capabilities of FAEs in the MSA task, the simplest backend is used in this study: connecting only a single-layer linear module as shown in Fig. \ref{fig:fig2_linear_pooling_model}.
With this setup, any ability to leverage long-term context is attributed solely to the FAE frontend, not the backend model.
During training, the backend takes 30 seconds of FAE features as input and estimates 30 seconds of MSA labels.
Since each FAE has a different frame rate, adaptive average pooling was applied to match the frame rate of the MSA labels.
The linear layer maps the $Z$-dimensional FAE features to an 8-dimensional embedding, with one used for boundary detection and seven for function prediction.
We used binary cross-entropy loss for boundary detection and 7-class cross-entropy loss for function prediction, respectively.
The sum of these losses was used for training.

\vspace{-2.5mm}
\subsection{Experimental Settings}
\label{sec:sec3_3_experimental_settings}
\vspace{-1mm}
We set the frame rate of the MSA labels to 2 Hz ($T=60$ in Fig. \ref{fig:fig2_linear_pooling_model}).
For training, we used a batch size ($B$) of 8, the AdamW optimizer~\cite{optimizer_adamw} with a weight decay of 0.01, a learning rate of 0.0001, a warmup period of 5 epochs, and cosine annealing for the remaining 95 epochs.
The model with the highest validation score was selected.
For post-processing, we applied the peak-picking algorithm~\cite{MSA_2014_ullrich} to detect boundaries, then chose the function labels with the highest average probability in each segment~\cite{MSA_2022_bytedance_icassp}.

\vspace{-2mm}
\section{Results and Discussion}
\label{sec:sec4_results_and_discussion}
\vspace{-2mm}
The results are shown in Table \ref{tab:table2_result_8fold}.

\vspace{-3mm}
\subsection{Results}
\label{sec:sec4_1_results}
\vspace{-1mm}
Our results suggest that FAEs trained with on \textit{music} using a \textit{long-context} Transformer are most effective for MSA to train FAEs for the MSA task, as supported by the following observations.

\smallskip
\noindent\textbf{Learning method.}
We found that most competitive models are all trained with MLM: MusicFM (MSD) is the most competitive model in this benchmark, as it achieves top-2 performance in all metrics;
AudioMAE (Zhong) achieved top-2 performance in boundary detection;
MERT (330M) showed strong performance in function prediction.
Compared to other SSL methods such as contrastive learning (MULE) and tokenization (EnCodec and DAC), MLM exhibits overwhelming success.
Supervised learning models (PANNs and PaSST) cannot match the MLM models trained on music data, which is consistent with prior research~\cite{AE_MERT,downstream_2025_sonido} that models pretrained with supervised learning cannot generalize well to unseen tasks.
Notably, fine-tuning the two AudioMAE models on the audio tagging task produces supervised models that outperform PANNs and PaSST, further emphasizing the importance of MLM pretraining.

\smallskip
\noindent\textbf{Model context length.}
Within MLM methods, MusicFM showed superior performance, which we hypothesize is a result of its longer context length: the 30-second context might have allowed MusicFM to better understand music structure compared to the 5-second context length in MERT and AudioMAE (Zhong).

\smallskip
\noindent\textbf{Training data.}
Although AudioSet contains a large amount of music, multiple short clips from the same music track are rare.
As a result, the model is never trained to distinguish those similar parts from the same track.
In contrast, when training on full-track or long-form music data, the model is presented with all parts from the same track.
Even though the model can only see a clip at a single training step, such training data encourages the model to distinguish between different parts of the same track, particularly for SSL methods that can learn features from data without annotations.
We hypothesize that full-track or long-form music data has been beneficial to MusicFM, MERT, and AudioMAE (Zhong).
On the other hand, AudioMAE (Huang), a weaker MLM model, was trained on AudioSet.

{
\setlength{\tabcolsep}{5pt}
\begin{table*}[t]
\centering
\footnotesize
\begin{tabular}{c|cc|cc|cc|cc}
\multirow{3}{*}{FAE}&\multicolumn{4}{c|}{Boundary detection}&\multicolumn{4}{c}{Function prediction}\\
&\multicolumn{2}{c|}{HR.5F}&\multicolumn{2}{c|}{HR3F}&\multicolumn{2}{c|}{PWF}&\multicolumn{2}{c}{ACC}\\
&No pooling&Pooling&No pooling&Pooling&No pooling&Pooling&No pooling&Pooling\\
\hline
\rowcolor{gray!20}\multicolumn{9}{l}{\it Self-supervised Learning: Masked Language Modeling (MLM)}\\
MusicFM (FMA)&51.22$\pm$1.12&41.39$\pm$1.58&58.80$\pm$0.84&59.19$\pm$1.23&\underline{66.35}$\pm$1.99&63.78$\pm$1.29&\underline{67.65}$\pm$1.59&63.14$\pm$0.97\\
MusicFM (MSD)&\textbf{54.19}$\pm$0.94&49.76$\pm$0.64&60.58$\pm$0.76&\underline{63.91}$\pm$1.18&\textbf{66.89}$\pm$1.52&64.66$\pm$1.14&\textbf{68.13}$\pm$1.84&64.44$\pm$1.55\\
MERT (95M)&42.94$\pm$0.85&42.23$\pm$1.93&52.25$\pm$0.56&60.99$\pm$1.27&62.44$\pm$1.22&63.40$\pm$1.33&62.58$\pm$1.16&62.01$\pm$1.27\\
MERT (330M)&45.39$\pm$1.01&40.63$\pm$1.88&54.46$\pm$0.99&57.72$\pm$1.96&64.16$\pm$1.30&64.17$\pm$1.37&63.77$\pm$1.37&62.30$\pm$1.46\\
AudioMAE (Huang)&36.57$\pm$1.17&36.95$\pm$1.18&51.09$\pm$1.54&58.11$\pm$1.09&60.33$\pm$0.88&64.58$\pm$1.49&59.65$\pm$1.24&63.07$\pm$1.93\\
AudioMAE (Zhong)&43.92$\pm$0.49&\underline{53.86}$\pm$1.07&59.26$\pm$0.80&\textbf{64.87}$\pm$0.98&62.85$\pm$1.24&64.06$\pm$1.71&60.99$\pm$1.23&61.33$\pm$2.02\\
\hline
\rowcolor{gray!20}\multicolumn{9}{l}{\it Self-supervised Learning: Contrastive Learning}\\
MULE&20.40$\pm$0.66&n/a&43.61$\pm$0.89&n/a&57.67$\pm$1.43&n/a&57.38$\pm$1.85&n/a\\
\hline
\rowcolor{gray!20}\multicolumn{9}{l}{\it Self-supervised Learning: Tokenization (Codec)}\\
EnCodec (24kHz / 24kbps)&23.98$\pm$0.73&19.25$\pm$1.47&43.00$\pm$0.72&31.81$\pm$0.85&53.94$\pm$1.13&52.87$\pm$1.14&48.52$\pm$1.39&45.77$\pm$2.14\\
EnCodec (48kHz / 24kbps)&23.42$\pm$1.09&19.67$\pm$1.40&42.60$\pm$0.94&34.77$\pm$1.33&54.42$\pm$0.96&53.44$\pm$0.96&52.40$\pm$1.59&44.63$\pm$1.75\\
DAC&23.33$\pm$1.06&19.10$\pm$0.93&42.73$\pm$0.81&39.63$\pm$0.96&54.79$\pm$1.35&55.06$\pm$0.83&50.34$\pm$1.76&50.21$\pm$1.42\\
\hline
\rowcolor{gray!20}\multicolumn{9}{l}{\it  Supervised Fine-tuning (Audio Tagging) after MLM}\\
AudioMAE (Huang)&44.26$\pm$0.70&38.41$\pm$1.34&57.23$\pm$0.89&59.14$\pm$0.42&63.30$\pm$1.61&63.95$\pm$1.92&63.25$\pm$1.70&63.14$\pm$1.64\\
AudioMAE (Zhong)&37.74$\pm$1.10&36.50$\pm$1.25&53.82$\pm$0.91&54.31$\pm$1.14&62.61$\pm$1.69&61.53$\pm$1.39&61.09$\pm$2.28&58.64$\pm$1.21\\
\hline
\rowcolor{gray!20}\multicolumn{9}{l}{\it Supervised Learning (Audio Tagging)}\\
PANNs&n/a&26.12$\pm$0.76&n/a&46.37$\pm$0.89&n/a&59.29$\pm$0.94&n/a&57.55$\pm$1.59\\
PaSST&28.94$\pm$1.08&22.00$\pm$0.96&45.52$\pm$0.87&44.06$\pm$1.20&59.28$\pm$1.08&58.39$\pm$1.56&57.61$\pm$1.10&55.80$\pm$1.94\\
\hline
\rowcolor{gray!20}\multicolumn{9}{l}{\it Supervised Learning \& Fine-tuning (Sound Event Detection)}\\
PANNs&28.73$\pm$0.83&23.89$\pm$0.72&53.22$\pm$0.72&46.73$\pm$0.79&60.01$\pm$1.29&57.60$\pm$1.23&58.45$\pm$1.24&54.90$\pm$1.06\\
\hline
\rowcolor{gray!20}\multicolumn{9}{l}{\it Cross-modal Contrastive Learning (Audio-text)}\\
CLAP (music-audioset)&n/a&29.21$\pm$0.96&n/a&46.60$\pm$1.30&n/a&60.36$\pm$1.08&n/a&58.56$\pm$1.21\\
CLAP (music-speech-audioset)&n/a&29.29$\pm$0.92&n/a&46.50$\pm$1.17&n/a&60.46$\pm$1.19&n/a&59.03$\pm$0.96\\
\hline
\rowcolor{gray!20}\multicolumn{9}{l}{\it Cross-modal Contrastive Learning (Audio-visual)}\\
OpenL3&38.33$\pm$1.24&22.65$\pm$0.86&50.24$\pm$0.95&44.48$\pm$1.20&60.30$\pm$1.88&60.15$\pm$1.05&58.09$\pm$2.40&58.45$\pm$1.23
\end{tabular}
\vspace{-1mm}
\caption{8-fold validation of linear probing on Harmonix. Metrics are explained in Sec.~\ref{sec:sec3_1_dataset_and_metrics}. Values are mean$\pm$standard deviation. The best and second best results are shown in bold and underlined, respectively.}
\label{tab:table2_result_8fold}
\vspace{-4mm}
\end{table*}
}

\vspace{-3mm}
\subsection{Discussion}
\label{sec:sec4_2_discussion}
\vspace{-1mm}
\noindent\textbf{Pooling.}
MLM models confirmed the effectiveness of pooling.
HR3F improved across all MLM models, and PWF improved with MERT, while the two AudioMAEs improved across all metrics.
Nevertheless, pooling can cause boundary position shifts, indicating that HR.5F may degrade for some methods.

\smallskip
\noindent\textbf{Feature frame rate.}
MULE, an SSL model using contrastive learning, showed low boundary detection performance, likely due to its coarse frame rate of 0.5 Hz and the resulting blurred boundaries, as shown in Fig. \ref{fig:fig1_FAE_feature}.
However, an excessively high frame rate, such as in the case of EnCodec and DAC, might also harm boundary detection.  
Among the best three models, MERT (75 Hz) and MusicFM (25 Hz) have a high frame rate, but AudioMAE (Zhong) has only 6.25 Hz, indicating a high frame rate is not essential for strong MSA performance.

\smallskip
\noindent\textbf{Semantic vs. acoustic SSL.}
The tokenizers (codecs) and MLM models are similar in their training objectives, as both require models to reconstruct the input signal. 
However, tokenizers performed very poorly in both boundary detection and function prediction.
We believe the following differences lead to the observed performance gap.
MLM encourages models to reconstruct original signals from a masked version without pursuing reconstruction details, resulting in semantic features.  
In contrast, tokenizers are trained to recover as many acoustic details as possible, rather than semantic ones.
In Fig. \ref{fig:fig1_FAE_feature}, numerous fine vertical stripes are visible for EnCodec and DAC when pooling is not applied, which is the characteristic of fine-grained details.  
Pooling reduces these vertical stripes, but blurs boundaries and further degrades performance in these models.
We also examined EnCodec with various bit rates, but observed no statistically meaningful performance difference.

\smallskip
\noindent\textbf{Architecture.}
PANNs (CNN) and PaSST (Transformer), as well as OpenL3 (CNN) and CLAP (Transformer), which share similar learning methods, achieve similar performance, indicating that model architecture is not the key factor for these methods.
However, for longer 30-second training data, as with MusicFM, an architecture with a long receptive field, such as Transformers, might be preferred.

\smallskip
\noindent\textbf{Models used as metrics to evaluate music generation quality.}
The following FAEs have been used as backbones in metrics such as FAD to evaluate the quality of music and audio generative models: MERT in~\cite{review_fad,review_mad}; EnCodec, DAC, and CLAP in~\cite{review_fad}; PANNs in~\cite{metrics_audioldm}; PaSST in~\cite{metrics_specmaskfoley}; and OpenL3, which has been used in long-form music evaluation~\cite{metrics_stable_audio_2}.
In this study, we found that, compared to most of the aforementioned models, MLM models, particularly MusicFM, exhibit a better understanding of music structure.
We encourage the community to revisit the validity of these metrics and hypothesize that MLM methods could serve as strong alternatives.
In a recent metric called MAD~\cite{review_mad}, MERT has been chosen as the backbone, which we believe is a positive sign.

\smallskip
\noindent\textbf{Future work.}
An important branch of SSL is auto-regression (AR), which is not included in this research, as we found that AR models, such as Jukebox~\cite{downstream_2022_jukebox} and SoniDo~\cite{downstream_2025_sonido}, are resource-intensive due to their more-than-5B-parameters.
It would be interesting to explore AR models at a scale similar to MLM models (less than 330M parameters) and evaluate their potential in the MSA task.
We also noticed that, instead of a linear probe, backends utilizing longer contexts have been shown to improve MSA performance~\cite{MSA_2025_bytedance}.
Moreover, different layers in Transformer-based FAEs tend to contain different features~\cite{downstream_2022_jukebox}.
Investigating features from different FAE layers and designing more powerful backends are promising directions for future work.

\vspace{-2mm}
\section{Conclusion}
\label{sec:sec5_conclusion}
\vspace{-1mm}
This study addressed the underexplored capabilities of foundational audio encoders (FAEs) for music structure analysis (MSA).
Through a comparative evaluation of 11 FAEs, we investigated key factors such as learning method, training data, and context length.
Our experimental results demonstrate that self-supervised masked language models trained on music data with a long context length achieve the strongest performance, a finding consistent with MSA's inherent need to capture long-term musical context.
Our findings provide clear guidance for future research on MSA and FAEs, suggesting that the strong FAEs identified in this study can also serve as improved backbones for metrics such as FAD to evaluate long-form music generation.

\clearpage
\bibliographystyle{IEEEbib}
{
\footnotesize
\bibliography{strings,refs}
}
\end{document}